\documentclass[apj]{emulateapj}

\newcommand{\parcsec}{\mbox{$\stackrel{\prime\prime}{\textstyle .}$}}

\usepackage{graphicx,amsmath,amssymb}

\slugcomment{ApJ, in press (accepted: 25 April 2005)}

\shorttitle{The Radio Afterglow and Host Galaxy of the Dark GRB\,020819}
\shortauthors{Jakobsson et al.}

\begin{document}


\title{THE RADIO AFTERGLOW AND HOST GALAXY OF THE DARK GRB\,020819}

\author{P.~Jakobsson,\altaffilmark{1,2} D.~A.~Frail,\altaffilmark{3}
D.~B.~Fox,\altaffilmark{4} D.-S.~Moon,\altaffilmark{4,5,6} 
P.~A.~Price,\altaffilmark{7,8} S.~R.~Kulkarni,\altaffilmark{4}
J.~P.~U.~Fynbo,\altaffilmark{1} J.~Hjorth,\altaffilmark{1} 
E.~Berger,\altaffilmark{9} R.~H.~McNaught,\altaffilmark{8}
and H.~Dahle\altaffilmark{10}}

\altaffiltext{1}{Niels Bohr Institute, University of Copenhagen, 
Juliane Maries Vej 30, 2100 Copenhagen, Denmark}

\altaffiltext{2}{Science Institute, University of Iceland, Dunhaga 3,
107 Reykja\-v\'{\i}k, Iceland}

\altaffiltext{3}{National Radio Astronomy Observatory, PO Box 0, 
Socorro, New Mexico 87801, USA}

\altaffiltext{4}{Division of Physics, Mathematics, and Astronomy, 
California Institute of Technology, MS 130-33, Pasadena, CA 91125, USA}

\altaffiltext{5}{Space Radiation Laboratory 220-47, California Institute 
of Technology, Pasadena, CA 91125, USA}

\altaffiltext{6}{Robert A. Millikan Fellow}

\altaffiltext{7}{Institute for Astronomy, University of Hawaii, 
2680 Woodlawn Drive, Honolulu, Hawaii 96822-1897, USA}

\altaffiltext{8}{Research School of Astronomy and Astrophysics, Mount 
Stromlo Observatory, Cotter Road, Weston Creek, Canberra, ACT 2611, 
Australia}

\altaffiltext{9}{Carnegie Observatories, 813 Santa Barbara Street,
Pasadena, CA 91101, USA}

\altaffiltext{10}{Institute of Theoretical Astrophysics, University of 
Oslo, P.O. Box 1029, Blindern, N-0315 Oslo, Norway}

\begin{abstract}
Of the fourteen gamma-ray bursts (GRBs) localized to better than 2\arcmin\
radius with the SXC on \emph{HETE-2}, only two lack optical
afterglow detections, and the high recovery rate among this sample has
been used to argue that the fraction of truly dark bursts is
$\sim$10\%.  While a large fraction of earlier dark bursts can be
explained by the failure of ground-based searches to reach appropriate
limiting magnitudes, suppression of the optical light of these SXC
dark bursts seems likely.  Here we report the discovery and
observation of the radio afterglow of GRB\,020819, an SXC dark burst,
which enables us to identify the likely host galaxy (probability of 
99.2\%) and hence the redshift ($z=0.41$) of the GRB.  The radio light 
curve is qualitatively similar to that of several other radio afterglows, 
and may include an early-time contribution from the emission of the
reverse shock.  The proposed host is a bright $R = 19.5$\,mag
barred spiral galaxy, with a faint $R\approx 24.0$\,mag ``blob'' of
emission, 3\arcsec\ from the galaxy core (16\,kpc in projection), that
is coincident with the radio afterglow.  Optical photometry of the
galaxy and blob, beginning 3~hours after the burst and extending over
more than 100 days, establishes strong upper limits to the optical
brightness of any afterglow or associated supernova.  Combining the
afterglow radio fluxes and our earliest $R$-band limit, we find that
the most likely afterglow model invokes a spherical expansion into a
constant-density (rather than stellar wind-like) external environment;
within the context of this model, a modest local extinction of
$A_V\approx 1$\,mag is sufficient to suppress the optical flux below
our limits.
\end{abstract}

\keywords{dust, extinction --- galaxies: high-redshift --- gamma rays: bursts}

\section{INTRODUCTION}
The host galaxies of gamma-ray bursts (GRBs) at low and intermediate 
redshifts are predominately sub-luminous, low-metallicity, and dust-poor 
galaxies with very blue colors \citep[e.g.,][]{fruchter,floch,fynboLya,lise}.
In particular, the three most nearby hosts (GRBs\,980425, 030329, and
031203) are most likely metal-poor dwarf galaxies similar to the Magellanic
Clouds, but with intense star formation \citep{fynbo2000,jens,prochaska}. 
The fact that GRB hosts are ubiquitously drawn from the population of 
faint, dust-poor galaxies, implies that either (i) such galaxies are 
responsible for the majority of star formation at all redshifts, or 
(ii) GRBs are preferentially caused by low-metallicity stars, or 
(iii) the current sample is observationally biased against GRBs 
occurring in brighter, high-metallicity hosts.
\par
Related to point (iii) above is the nature of dark bursts, whose
definition has been quite vague in the literature. It was commonly
used for those GRBs that were not accompanied by detections of 
optical afterglows (OAs), irrespective of how inefficient the 
search was. It is important to reveal if the 
fraction of dark bursts \citep[as high as 60\%--70\%, e.g.,][]{lazzati} 
is predominantly the result of inadequate observing strategies, or if 
dust and high redshift play a major role. Evidence has been mounting 
\citep{fynbo2001,edoDARK} that the former is the primary reason for such 
a high fraction. Recently, \citet{palli} proposed a revised definition of 
dark bursts, as those bursts that are optically subluminous with respect to 
the fireball model, i.e., which have an optical-to-X-ray spectral index 
$\beta_{\mathrm{OX}} < 0.5$. Applying this definition to a set of 52 
bursts, only 10\% have optical limits deep enough to establish them
as dark bursts. \citet{rol} obtained a comparable dark burst fraction
using a more detailed analysis, albeit with a smaller sample. 
\par
\begin{deluxetable*}{@{}lrlclcccc@{}}
\tablecaption{Optical/NIR Observations of GRB\,020819 \label{optical.tab}}
\tablehead{
Epoch                                          &
$\Delta t$\tablenotemark{a}                    &
Telescope/                                     &
Seeing                                         &
Exposure Time                                  &
Filter                                         &
\multicolumn{2}{c}{Magnitude\tablenotemark{b}} &
Limiting\tablenotemark{c}                      \\
                 & 
[days]           &
Instrument       &
[arcsec]         &
\hspace{1cm}[s]  &
                 &
Galaxy           &
Blob             &
Magnitude  \hspace{-2mm}
}
\startdata
2002 Aug 19.76 & 0.14   & 1.0-m SSO  & 2.5 & $3 \times 300$ & $R$ 
& \multicolumn{2}{c}{$19.62 \pm 0.15$} & 21.5 \\

2002 Aug 20.34 & 0.72   & P60        & 1.6 & $4 \times 600$ & $R$ 
& \multicolumn{2}{c}{$19.60 \pm 0.11$} & 22.1 \\

2002 Aug 26.53 & 6.91   & Keck/NIRC\tablenotemark{d,e} & --- & 
$9 \times 60$ & $K$ & \multicolumn{2}{c}{---} & ---  \\

2002 Sep 03.47 & 14.85  & Keck/ESI      & 0.6 & $7 \times 600$ & $B$ 
& $21.90 \pm 0.50$ & $26.10 \pm 0.50$ & 26.9 \\

2002 Sep 03.51 & 14.89  & Keck/ESI      & 0.6 & $4 \times 600$ & $R$ 
& $19.48 \pm 0.12$ & $23.99 \pm 0.13$ & 26.9 \\

2002 Sep 10.09 & 21.47  & NOT/ALFOSC    & 1.1 & $5 \times 600$ & $R$ 
& \multicolumn{2}{c}{$19.34 \pm 0.06$} & 25.3 \\

2002 Sep 15.17 & 26.55  & VLT/FORS2     & 0.7 & $22 \times 300$ & $R$ 
& $19.45 \pm 0.02$ & $23.98 \pm 0.06$ & 25.8 \\

2002 Oct 13.37 & 54.75  & Keck/NIRC\tablenotemark{d} & --- &   
$27 \times 60$ & $K$ & $16.80 \pm 0.20$ & $20.80 \pm 0.30$ & ---  \\

2003 Jan 01.24 & 134.62 & Keck/ESI      & 0.7 & $2 \times (300+600)$ & $R$ 
& $19.43 \pm 0.10$ & $23.76 \pm 0.15$ & 26.6 \hspace{-7mm}

\enddata

\tablecomments{The magnitudes reported here are the results of
               aperture photometry (see main text for details).
               Correction for Galactic extinction has been
               applied to the photometry \citep{schlegel}.}

\tablenotetext{a}{Measured in days after 2002 August 19.623 UT.}
\tablenotetext{b}{The host galaxy and blob were not resolved in the
                  SSO, P60, and NOT observations.}
\tablenotetext{c}{Limiting magnitudes are $2 \sigma$ in a circular 
                  aperture with a radius equal to the seeing.}
\tablenotetext{d}{There were no stars in the Keck/NIRC field of
                  view to estimate the seeing and limiting magnitude.}
\tablenotetext{e}{No standard star observations were available for
                  calibration of this epoch.}

\end{deluxetable*}
In this paper we present the detection of the radio afterglow of
GRB\,020819, and propose that a bright spiral galaxy ($z = 0.41$) 
coincident with the radio position is the host. We explore the nature 
of the GRB external environment, by applying the fireball model on the 
afterglow radio flux. When combined with an early optical $R$-band limit, 
this reveals if host extinction needs to be invoked. We also search
for a signature of a supernova (SN). Finally, we discuss GRB\,020819 in 
the context of rapid localizations and follow-up of GRBs, and its 
implications for dark bursts. We adopt a cosmology where the Hubble 
para\-meter is $H_0 = 70$\,km\,s$^{-1}$\,Mpc$^{-1}$, $\Omega_{\mathrm{m}} 
= 0.3$, and $\Omega_{\Lambda} = 0.7$. For these parameters, a redshift of 
0.41 corresponds to a luminosity distance of 2.24\,Gpc and a distance 
modulus of 41.7. One arcsecond is equivalent to 5.45 proper kiloparsecs, 
and the lookback time is 4.36\,Gyr.
\section{OBSERVATIONS}
GRB\,020819 was detected by the French Gamma Telescope
(FREGATE), Wide Field X-ray Monitor (WXM) and Soft X-ray Camera (SXC)
on-board the \emph{High Energy Transient Explorer 2} (\emph{HETE-2}) 
satellite on 2002 August 19.623 UT. A $130\arcsec$ radius SXC error 
circle was circulated 3 hours after the burst \citep{vanderspek}. A few 
days later its location was refined and the radius reduced to $64\arcsec$, 
lying fully within the original error circle \citep{crew}.
\par
We observed the $130\arcsec$ radius SXC error circle with the 
Siding Spring Observatory (SSO) 1.0-m telescope in Australia at
$\Delta t = 3.1$\,hr \citep{price}, where $\Delta t$ is the time from 
the onset of the burst. At $\Delta t = 17.3$\,hr we imaged the
field again with the \mbox{1.5-m} telescope (P60) at Palomar 
\citep{price2}. In neither case was an OA detected. We continued to 
monitor the field in the optical/NIR during the following weeks and 
months with the Echellette Spectrograph and Imager (ESI) and the Near 
Infrared Camera (NIRC) on Keck, the Andalucia Faint Object Spectrograph 
and Camera (ALFOSC) on the Nordic Optical Telescope (NOT), and the 
\mbox{FOcal} Reducer and low dispersion Spectrograph (FORS2) on the 
Very Large Telescope (VLT). The journal of our observations is 
listed in Table~\ref{optical.tab}.
\par
Although a rapid and deep follow-up was performed, no optical/NIR 
afterglow has been reported for GRB\,020819. \citet{levan} obtained a
deep optical limit ($R = 22.2$\,mag) at $\Delta t = 9$\,hr, covering
the original SXC error circle. During the same epoch, \citet{klose}
performed a NIR follow-up observation, with a limiting magnitude of 
$K = 19.5$\,mag. Unfortunately, their analysis and afterglow search was 
only applied to the refined $64\arcsec$ SXC error circle (see discussion 
in Sect.~\ref{radio.sec}).
\par
\begin{deluxetable}{@{}lrcr@{}}
\tablecaption{VLA Observations of GRB\,020819 \label{radio.tab}}
\tablewidth{210pt}
\tablehead{
Epoch                       &
$\Delta t$\tablenotemark{a} &
$\nu_{\mathrm{obs}}$        & 
$S \pm \sigma$              \\
          & 
[days]    &
[GHz]     & 
[$\mu$Jy]  
}

\startdata

2002 Aug 21.37 & 1.75   & 8.46 & $315 \pm 18$ \\
2002 Aug 23.34 & 3.72   & 8.46 & $176 \pm 19$ \\
2002 Aug 23.36 & 3.74   & 1.43 & $39  \pm 60$ \\
2002 Aug 29.32 & 9.70   & 8.46 & $264 \pm 30$ \\
2002 Aug 29.34 & 9.72   & 4.86 & $224 \pm 34$ \\
2002 Sep 10.27 & 21.65  & 8.46 & $161 \pm 24$ \\
2002 Sep 18.26 & 29.64  & 8.46 & $109 \pm 20$ \\
2002 Oct 03.17 & 44.55  & 8.46 & $43  \pm 31$ \\
2002 Dec 24.07 & 126.45 & 8.46 & $79  \pm 25$ \\ 
2002 Dec 28.02 & 130.40 & 8.46 & $32  \pm 26$ \\
2003 Jan 10.06 & 143.44 & 8.46 & $62  \pm 25$ \\ 
2003 Jan 23.90 & 157.28 & 8.46 & $6   \pm 14$ \hspace{-3mm}
\enddata

\tablenotetext{a}{Measured in days after 2002 August 19.623 UT.}

\end{deluxetable}
We discovered \citep{frail} a variable radio source at 8.46\,GHz with the 
Very Large Array (VLA), declining from a peak of 315\,$\mu$Jy at 
$\Delta t = 1.75$ days to a level of non-detection at $\Delta t = 
157$ days. A log of the GRB\,020819 radio observations is given in 
Table~\ref{radio.tab}. This source was located at $\alpha$(J2000) = 
$23^{\mathrm{h}}27^{\mathrm{m}}19\fs475$ and $\delta$(J2000) = 
$06\arcdeg15\arcmin55\farcs95$, with an error of $0\farcs5$ in each
coordinate. The radio transient is located only $3\arcsec$ away from 
the center of a bright barred spiral galaxy. In addition, the transient is 
superimposed on a faint blob of light. A Keck/ESI image covering the 
surrounding field is displayed in Fig.~\ref{find.fig}. 
\section{RESULTS}
\subsection{Optical/NIR}
The optical data were reduced using standard techniques for de-biasing 
and flat-fielding. In order to determine the photometric properties 
of the potential host galaxy and blob, we used aperture photometry.
A circular aperture with a fixed radius of $4\parcsec5$ was used
for all the different telescopes/instruments to obtain the total flux 
from the galaxy. To verify the results, we used SExtractor \citep{sex} 
to obtain its total magnitudes (\texttt{mag\_auto}). The photometry of
the nearby blob was also carried out using aperture photometry, with
a circular aperture with a fixed radius of $0\parcsec8$. Only the
VLT/FORS2 epoch had a simultaneous $R$-band observation of a \citet{landolt}
standard field. We used this to calibrate secondary standards for the
GRB\,020819 field, which we in turn used to transform the relative 
optical magnitudes to the Cousins photometric system. For the 
$B$-band Keck epoch there was only one faint non-saturated calibrated 
star in field from the Digitized Second Palomar Observatory Sky 
Survey (DPOSS). The resulting calibrated magnitude is therefore
quite uncertain. We note that DPOSS uses the Gunn photometric system, 
which we converted to the Johnson photometric system 
\citep[e.g.,][]{jorgensen}. 
\begin{figure}
\centering
\fbox{\includegraphics[width=8.33cm]{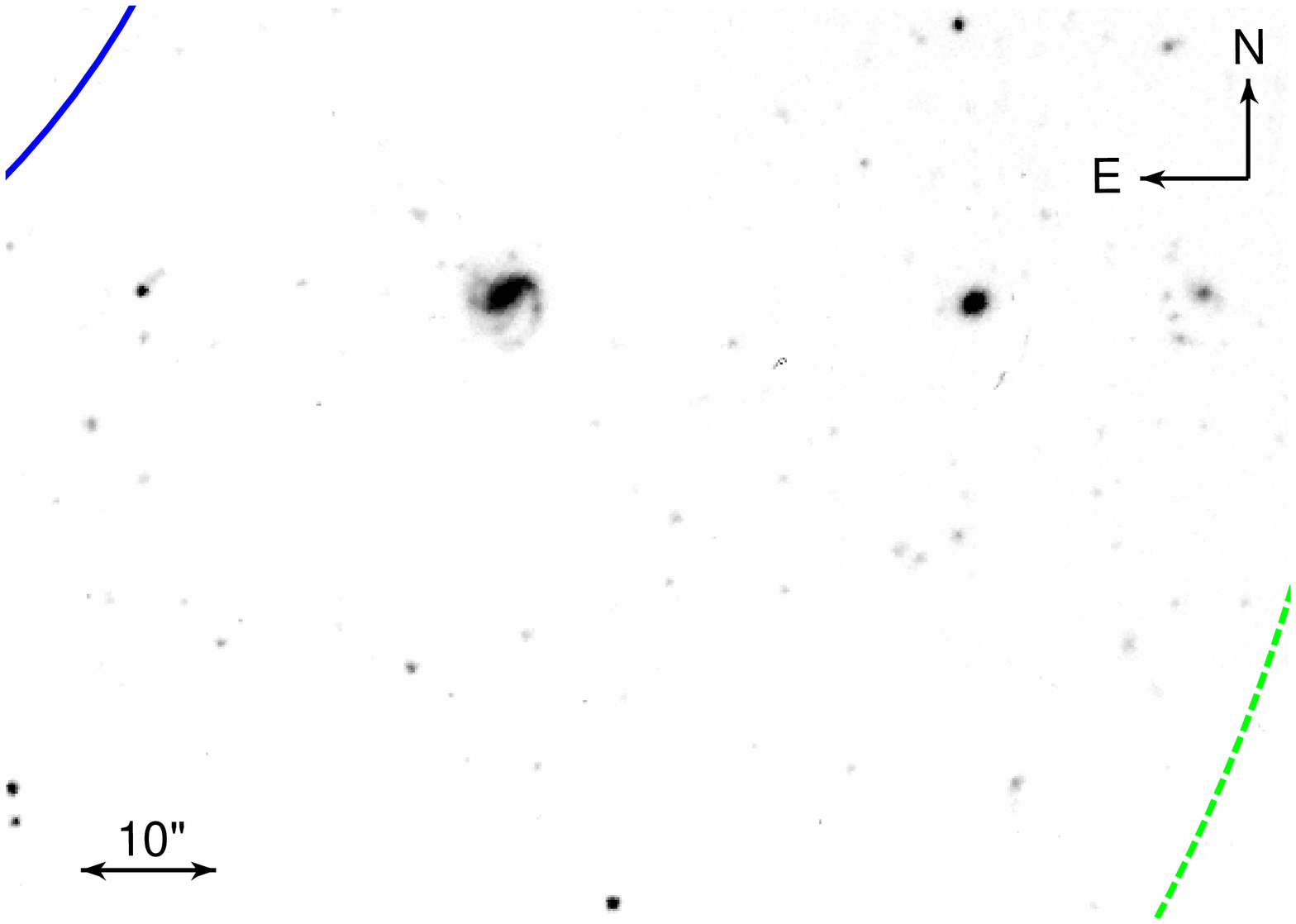}}
\fbox{\includegraphics[width=8.33cm]{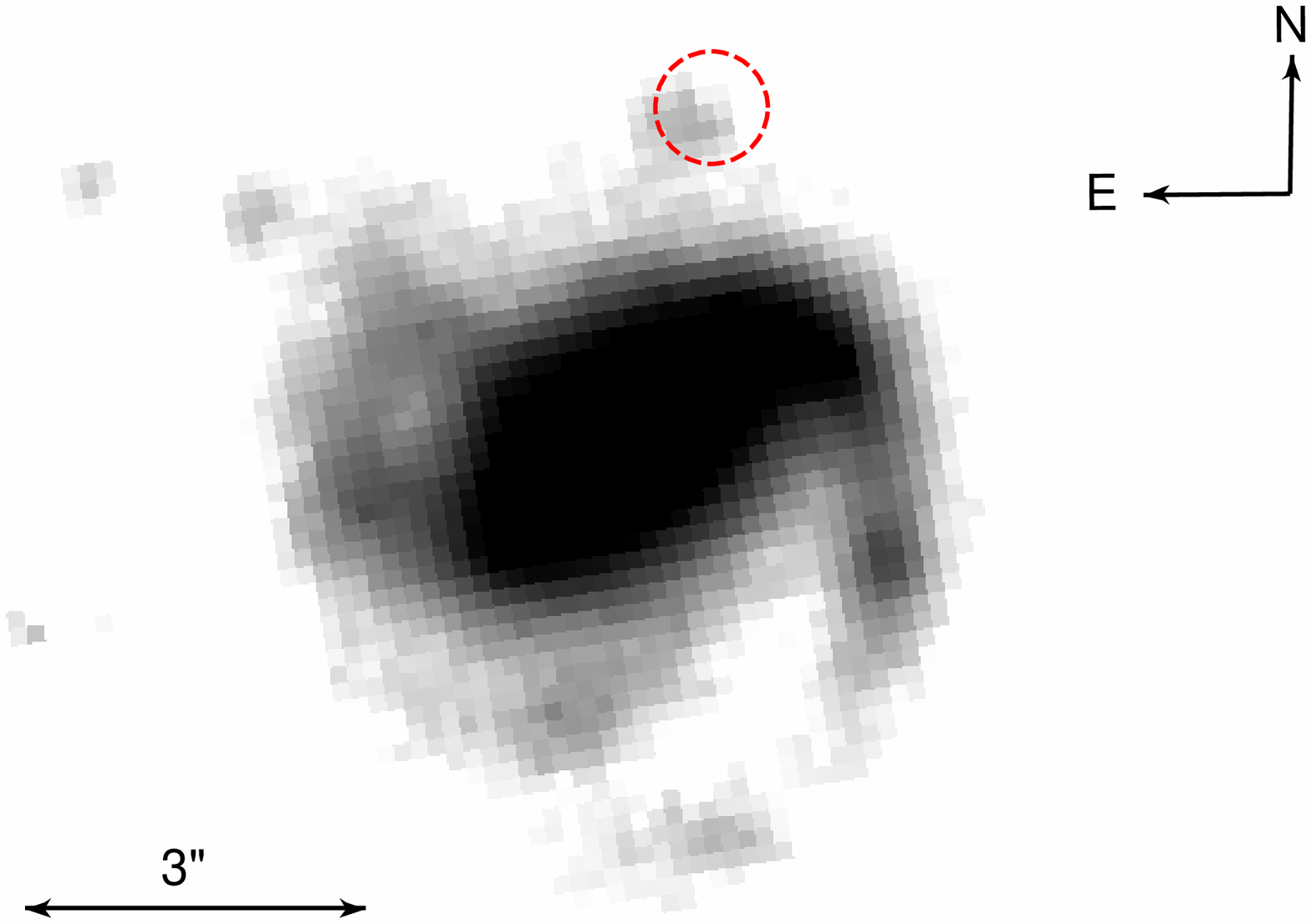}}
\caption{A  2400\,s $R$-band Keck/ESI image of the neighborhood of 
    the GRB\,020819 radio afterglow, obtained approximately 15 days 
    after the burst occurred. \emph{Top:} the original $130\arcsec$
    radius error circle is indicated by the dashed line, while
    the refined one is shown with a continuous line. \emph{Bottom:} 
    a zoom in on the position of the radio afterglow, marked with
    a dashed circle with a radius of $0\farcs5$. A bright barred 
    spiral galaxy is located around $3\arcsec$ from the afterglow,
    which is superpositioned on a faint lightsource. The image is 
    shown with a logarithmic intensity scale to highlight the
    features of the nearby blob.}
\label{find.fig}
\end{figure}
\par
The NIR images were dark-subtracted and flat-fielded using the
object-masked average of proximate science frames.  They were
registered against each other by reference to the centroid position of
the single bright object, the galaxy core, and co-added.  A rough WCS
was put on the images using the telescope pointing and instrument
rotation information as encoded in the image headers; this was later
refined by reference to larger-scale images of the field. Galaxy 
and blob fluxes were obtained with aperture photometry. For the 
second $K$-band epoch, we used an observation of a 2MASS star to 
transform the magnitudes to the standard system. Unfortunately, no 
standard stars were available for the first $K$-band epoch.
\par
The results of the optical/NIR photometry are listed in 
Table~\ref{optical.tab}. We have an early, relatively deep
limit of $R > 21.5$\,mag at $\Delta t = 3.4$\,hr. This is fainter than
the entire sample of afterglows detected and studied in the past 
seven years at a similar epoch, excluding the recent GRB\,050126 
\citep[see figure 5 in][]{edo}. We note, however, that much deeper
optical limits have been obtained at a similar epoch, e.g., 
GRB\,970828 \citep{groot}. Whether the GRB\,020819 optical faintness is 
due to high redshift, host extinction, or simply an intrinsically faint 
GRB, is a question best answered in synergy with the radio observations.
\par
The bright barred spiral galaxy and the faint blob are clearly
resolved in the Keck and VLT images. After correcting for foreground 
(Galactic) extinction using the reddening maps of \citet{schlegel},
the galaxy has $R - K = 2.7 \pm 0.2$\,mag, while the nearby blob has 
$R - K = 3.2 \pm 0.3$\,mag. To examine if a NIR afterglow contributes 
flux to the blob in the first $K$-band epoch ($\Delta t = 6.9$ days), 
we performed an image subtraction using the second $K$-band epoch. We 
subtracted out the galaxy, and looked for residual emission from the 
blob. But due to the lack of stars in the field of view, the validity 
of the subtraction could not be confirmed. The error in the galaxy 
subtraction remained larger than any residual emission from the blob. 
\subsection{The Radio Light Curve}
\label{radio.sec}
All observations were performed using the VLA in its standard
continuum mode. At each frequency, the full 100\,MHz bandwidth
was obtained in two adjacent 50\,MHz bands. The flux density scale
was tied to the extragalactic source 3C\,147, while the array
phase was monitored by switching between the GRB and a VLA phase
calibrator J2320+052 (J2330+110) at 8.46\,GHz (1.43\,GHz).
Data calibration and imaging were carried out with Astronomical 
Image Processing System software package.
\par
\begin{figure}[!t]
\epsscale{1.22}
\plotone{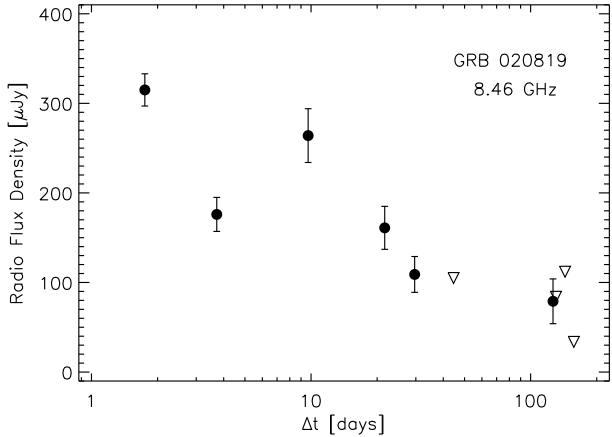}
\caption{An 8.46 GHz light curve from the VLA of the radio afterglow of 
    GRB\,020819. Detections at each epoch are indicated by filled circles.
    Non-detections, shown with open triangles, are defined as the peak 
    brightness at the location of the afterglow plus two times the rms 
    noise in the image.}
\label{radio.fig}
\end{figure}

In an image from $\Delta t = 1.75$~days, we found an unresolved
source at high significance (18$\sigma$). Although the transient 
is $98\arcsec$ away from the center of the revised $64\arcsec$ SXC 
error circle, its position and radius have been revised again 
\citep{villasenor}. It now includes the
radio transient, lending support to the hypothesis that the transient 
is in fact the GRB afterglow. Its light curve is presented in 
Fig.~\ref{radio.fig}. It is qualitatively similar to previous
radio afterglows, where the first data point is most likely
due to a strong reverse shock \citep[see a schematic light curve
in][]{frailConf}, GRB\,990123 being the best-known example
\citep{reverse}. 
\par
There is a sign of a late-time flattening in the radio light
curve. Due to the relatively deep upper limit at $\Delta t 
\approx 157$~days, it is unlikely that the final radio detection 
at $\Delta t \approx  126$~days originates from the host galaxy.
A more probable explanation is an episodic injection of energy,
where slower moving shells of ejecta catch up with the decelerating
main shock and re-energize it. This refreshed shock model has recently 
been invoked to explain the optical, X-ray, and radio behavior of 
GRB~021004 \citep{gulli}. We note that the flux limit of
$34\,\mu$Jy at $\Delta t \approx 157$~days implies an upper limit to
the host star-formation rate of approximately 
$40\,M_{\odot}\,\textrm{yr}^{-1}$ \citep[see figure 1 in][]{edoSFR}.
\par
The qualitative behavior of other radio light curves can be
summarized as follows \citep[e.g., figures 1 and 2 in][]{sari}.
The flux at a given radio frequency ($\nu_{\mathrm{R}}$) rises
as a power law $F_{\mathrm{R}}(t) \propto t^{\alpha_{\mathrm{r}}}$,
reaches a broad maximum at $F_{\mathrm{m}}$, and then decays as 
$F_{\mathrm{R}}(t) \propto t^{\alpha_{\mathrm{d}}}$. This is due
to the evolution of the characteristic break frequencies in 
the broadband synchrotron spectrum, where e.g., the epoch of
peak flux is frequency dependent ($t_{\mathrm{m}}(\nu_{\mathrm{R}})$).
\par
We have carried out a fit to this model, but given the paucity of our 
data, we fixed $t_{\mathrm{m}}(8.46\,\mathrm{GHz}) = 10$ days, a 
typical and reasonable (Fig.~\ref{radio.fig}) value. In addition,
we have omitted the first and final detection (see discussion above).
From the formal best fit we find $\alpha_{\mathrm{r}} = 0.44 \pm 0.16$, 
$\alpha_{\mathrm{d}} = -0.78 \pm 0.17$, with 
$F_{\mathrm{m}}(8.46\,\mathrm{GHz}) = 271 \pm 31$\,$\mu$Jy 
($\chi^2_{1} = 0.42$, where $\chi^2_{\mathrm{dof}}= \chi^2/
\mathrm{degree~of~freedom}$, is the reduced $\chi^2$ of the fit).
Shifting $t_{\mathrm{m}}$ a few days on either side of our value
does not substantially change the fitted parameters.
\section{DISCUSSION}
\label{dis.sec}
\subsection{The Host Galaxy}
The question whether a high redshift of the proposed host galaxy 
contributes significantly to the optical faintness of the
afterglow is easily answered by the results of Berger et al.
(in preparation). They measure a redshift of $z = 0.41$ for the
host\footnote[1]{Unfortunately, there is no conclusive redshift 
reported for the blob.}, the eighth closest thus far observed. As a 
comparison, the GRB\,011121 afterglow ($z = 0.36$) had a radio flux of 
610\,$\mu$Jy at 7 days \citep{price011121} and $R \approx 18$\,mag
at 0.4 days \citep{garnavich,germany}. These numbers indicate that 
the galaxy is a promising host candidate, although it needs to be
explored if the radio afterglow and the $R$-band limit are 
consistent with the fireball model (see Sect.~\ref{model.sec}).
\par
We have estimated the probability that the assigned host is a
chance superposition and not physically related to the GRB,
following \citet{bloom}. As the GRB localization is good, but
the position is outside the majority of the light of the spiral,
we have calculated the effective radius \citep[see 
equations~1--3 in][]{bloom} as $(R_0^2 + 
4 R_{\mathrm{half}}^2)^{1/2}$. Here $R_0 = 3\farcs0 \pm 0\farcs1$
is the radial separation between the GRB and the presumed host,
and $R_{\mathrm{half}} = 1\farcs35 \pm 0\farcs05$ is the half-light
radius. This, combined with the galaxy magnitude, results in a chance 
superposition probability of only 0.8\%. 
\par
The $R$$-$$K$ colors of galaxies provide crucial information on
the importance of old stellar populations (as traced by the
NIR emission), relative to the contribution of young stars
dominating the optical light. Blue objects ($R - K \sim 2$--3\,mag)
are a sign of unobscured star-forming galaxies, while old
ellipticals and dust-enshrouded starburst galaxies have large
$R$$-$$K$ colors. An \mbox{$R$$-$$K = 2.7$\,mag} is a bit higher than the 
mean value, but within the range spanned by the sample of 15 host 
galaxies explored by \citet{floch}. A redshift of $0.41$ places the host 
on a curve in figure 2 in \citet{floch} matching a type Scd galaxy, roughly 
consistent with the morphology seen in Fig.~\ref{find.fig}. In general,
Sc/Scd spirals display resolved \ion{H}{2} regions (the blob being
a likely candidate) and contain a larger fraction of dust and 
gas, which might explain the optical faintness of the afterglow. We
note that all the spiral hosts observed so far are nearly face-on
(GRBs\,980425, 990705, 020819). Although this sample is small, it
resembles SN-selected spirals which are more often 
observed face-on, most likely due to dust-bias \citep[e.g.,][]{cappellaro}.
\subsection{Afterglow Modeling}
\label{model.sec}
The nature of the ambient medium in which the GRB originated can
be probed with the parameters $F_{\mathrm{m}}$, $t_{\mathrm{m}}$,
and $\alpha_{\mathrm{d}}$. In addition, they can be used to predict 
the $R$-band magnitude at $\Delta t = 0.14$~days. We consider three
afterglow models: (i) an isotropic expansion into a homogeneous medium 
\citep{sari}, (ii) an isotropic expansion into a wind-stratified medium 
\citep{wind}, and (iii) a collimated expansion (jet) into a homogeneous 
or wind stratified medium \citep{jet}.
\par
For (i), the fits in Sect.~\ref{radio.sec}, predict
that the optical flux would have reached the maximum value
of $F_{\mathrm{m}}$ (corresponding to $R = 17.6$\,mag) at an epoch
$t_{\mathrm{m}}(\mathrm{opt}) = t_{\mathrm{m}}(8.46\,\mathrm{GHz}) 
\times (\nu_{\mathrm{R}}/\nu_{\mathrm{opt}})^{2/3} \approx 600$\,s.
The expected flux at later times would then drop by
$2.5 \alpha_{\mathrm{d}} \log(\Delta t/t_{\mathrm{m}}(\mathrm{opt}))$
magnitudes, which for $\Delta t = 0.14$~days corresponds to 2.5\,mag.
In this framework, we would thus expect to detect an OA
with $R = 20.1 \pm 0.6$\,mag at the epoch of the first observation.
Comparing to the upper limit of $R > 21.5$\,mag, only a modest amount
of extinction must be invoked. The electron energy power-law index
in this model is predicted to be $p = 1 - 4/3\alpha_{\mathrm{d}} 
= 2.04 \pm 0.23$, consistent with what is usually found 
\citep[$\bar{p} = 2.26$:][]{piro-p}
\par
As in (i), the optical flux in (ii) reaches $F_{\mathrm{m}}$ at
an epoch $t_{\mathrm{m}}(\mathrm{opt}) \approx 600$\,s. On the other hand,
since the afterglow emission is weakened as the relativistic blast
wave ploughs into ambient material with decreasing density, 
$F_{\mathrm{m}}$ is larger when extrapolated into the optical band,
with $F_{\mathrm{m}} \propto t^{-1/2}$. This results in an
afterglow with $R \approx 13.6$\,mag at 600\,s, decaying to 
$R = 16.1 \pm 0.6$\,mag at $\Delta t = 0.14$~days. This model would 
thus require a large host extinction. However, this model predicts 
$p = (1 - 4 \alpha_{\mathrm{d}})/3 = 1.37 \pm 0.23$. Although such
a low value for $p$ has previously been observed in an
afterglow \citep[e.g., GRB~010222:][]{galama03}, we are not in 
favor of this model.
\par
Finally, in model (iii), the radio emission (above the synchrotron
self-absorption frequency) is supposed to rise as $t^{1/2}$, reach
a maximum, and then slowly decline as $t^{-1/3}$ for 
$t > t_{\mathrm{jet}}$, where $t_{\mathrm{jet}}$ is the epoch when
the bulk Lorentz factor of the ejecta becomes comparable to the
inverse of the opening angle of the jet. The radio flux is then
expected to drop sharply as $t^{-p}$, when the maximum frequency falls 
below the observing frequency. This behavior has been observed for,
e.g., GRB\,990510 \citep{harrison}. As seen in Fig.~\ref{radio.fig},
the radio light curve does not show this kind of evolution. Even
if we assume $t_{\mathrm{jet}} < t_{\mathrm{m}} \approx 10$~days,
the model predicts $p = - \alpha_{\mathrm{d}} = 0.78$, a value not
considered relevant in afterglow models.
\subsection{A Limit on a Supernova Contribution}
An excess of optical flux at $\Delta t \sim 15 (1+z)$~days, superposed 
on a typical light curve power-law decay, is commonly attributed to the 
rise of a SN component \citep[e.g.,][]{galama,bloomSN1,bloomSN2}. 
Motivated by this, we timed our $R$-band observations ($\Delta t \approx 
15$, 21, 27~days) for a SN search. To accommodate the spread in the 
observed properties of Type Ibc SNe, we have considered the highly 
luminous SN\,1998bw associated with GRB\,980425 \citep{galama}, and
SN\,2002ap \citep{foley} which was a factor of $\sim$10 times less
luminous at peak time in the restframe $B$-band.
\par
At $z = 0.41$ the observed $R$-band corresponds to restframe $B$-band.
Thus, we have transformed the $B$-band magnitude at peak time of
the aforementioned SNe to AB magnitudes \citep{fukugita}, applied
the distance modulus, and the $2.5 \log (1+z)$ term 
\citep[see, e.g.,][]{vanDokkum}. If GRB\,020819 is associated with a
1998bw-like SN, we would expect a peak magnitude of $R = 22.3$\,mag at
$\Delta t \approx 20$~days. For a 2002ap-like SN the corresponding 
magnitude is $R = 24.8$\,mag.
\par
In Table~\ref{optical.tab}, the blob magnitude at $\Delta t \approx 
135$~days, when a SN and an OA contribution should be negligible, is 
$R = 23.76 \pm 0.15$\,mag. Adding the flux from a SN at $\Delta t \approx
20$~days would result in $R \sim 22.0$\,mag ($R \sim 23.4$\,mag) for 
SN\,1998bw (SN\,2002ap). Since we measure the blob magnitude to be 
$R = 23.99 \pm 0.13$\,mag and $R = 23.98 \pm 0.06$\,mag at $\Delta t 
\approx 15$~days and $\Delta t \approx 27$~days, respectively, we do
not detect the signature of a SN. This non-detection of a rising
SN component may be explained by a modest host extinction, in 
agreement with the afterglow data (see Sect.\ref{model.sec}). A 
2002ap-like SN extinguished by 1.4\,mag would remain undetected, 
alleviating the requirement of a very subluminous SN 
\citep[see also][]{zeh,soderberg}.
\begin{deluxetable*}{@{}lcrrcrrcll@{}}
\tablecaption{SXC Localized \emph{HETE-2} Gamma-Ray Bursts \label{sxc.tab}}
\tablewidth{0pt}
\tablehead{
GRB &
$\Delta t_{\mathrm{SXC}}$ &
$\Delta t_{\mathrm{OA}}$ & 
Mag($\Delta t_{\mathrm{OA}}$) & 
$\Delta t_1$ & 
Mag($\Delta t_1$) & 
$R$(1 day) &
$z$ &
SXC circle change? &
References \\
 &
[hours] &
[hours] & 
 & 
[hours] & 
 & 
 &
 &
 &
}

\startdata

020813        & 3.1  & 1.90 & $B \sim 18.6$ & 0.30 &
$CR > 16.0$   & 20.5 & 1.26 & Yes, OA outside & (1) (2) (3) (4) \\

020819        & 3.0  & --- & --- \hspace{4mm} & 2.98 &  
$R > 18.9$    & $>22.0$ & --- & Yes, RA inside & (5) (6) \\

021004        & 2.6  & 0.05 & $R = 15.5$ & 0.05 &   
$R = 15.5$    & 19.5 & 2.34 & Yes, OA inside & (7) (8) \\

021211        & 2.2  & 0.02 & $R = 14.1$ & 0.02 &   
$R = 14.1$    & 23.0 & 1.01 & No & (9) (10) \\

030115        & 1.4  & 2.05 & $R \sim 21.5$ & 0.01 &   
$R > 10.0$    & --- & --- & No & (11) (12) \\

030226        & 2.0  & 2.63 & $R \sim 17.5$ & 0.00 &   
$R > 11.5$    & 20.6 & 1.99 & No & (13) (14) \\

030328        & 0.9  & 1.28 & $R \sim 18.0$ & 1.28 &   
$R \sim 18.0$ & 21.5 & 1.52 & Yes, OA inside & (15) (16) \\

030329        & 1.2  & 1.56 & $R = 12.6$ & 1.56 &   
$R = 12.6$    & 16.0 & 0.17 & Yes, OA outside & (17) \\

030429        & 1.9  & 3.48 & $R = 19.7$ & 1.87 &   
$R > 18.7$    & 21.0 & 2.66 & Yes, OA outside & (18) (19) \\

030528        & 1.8 & 16.04 & $J = 20.6$ & 0.04 &   
$CR > 15.8$   & --- & ---   & Yes, OA inside & (20) (21) \\

030723        & 7.2 & 4.17 & $R = 20.9$ & 0.02 &   
$CR > 18.1$   & 21.1 & --- & Yes, OA inside & (22) (23) \\

040511        & 2.2 & 12.20 & $J \sim 19.0$ & 2.18 &
$CR > 17.2$   & --- & ---   & No & (24) (25) \\

041211        & 2.4 & --- & --- \hspace{4mm} & 0.01 &
$B > 11.0$    & --- & ---   & No & (26) \\

050408        & 1.2  & 2.40 & $R = 20.4$ & 0.01 &
$CR > 14.7$   & 22.3 & 1.24 & Yes, OA inside & (27) (28) (29)

\enddata

\tablerefs{(1) \citet{fox}; (2) \citet{rhoads}; (3) \citet{rykoff};
(4) \citet{gladders}; (5) \citet{urata}; (6) \citet{levan};
(7) \citet{foxNature}; (8) \citet{holland}; (9) \citet{wozniak};
(10) \citet{holland021211}; (11) \citet{masetti}; (12) \citet{alberto}; 
(13) \citet{fox030226}; (14) \citet{klose030226}; 
(15) \citet{peterson}; (16) \citet{andersen}; 
(17) \citet{priceNature}; (18) \citet{smith};
(19) \citet{palli429}; (20) \citet{uemura}; (21) \citet{rau}; 
(22) \citet{smithquimby}; (23) \citet{fynbo};
(24) \citet{smith511}; (25) \citet{fox511};
(26) \citet{sasaki}; (27) \citet{smith2005};
(28) \citet{misra}; (29) \citet{edoGCN}.
}

\tablecomments{A list of all GRBs localized to better than 
2\arcmin\ radius with the SXC, and where the first optical observation 
was performed within approximately 3~hours of the high energy event. Here
$\Delta t_{\mathrm{SXC}}$ is the time from the onset of the burst when
the $\sim$$2\arcmin$ SXC error circle radius was first made available.
$\Delta t_{\mathrm{OA}}$ is the time after the burst when the OA was
detected. $\Delta t_{\mathrm{1}}$ is the time after the GRB
occurred when the first optical observation was performed. In the
magnitude columns, 'CR' denotes filterless observations. In the 
penultimate column, 'RA' refers to the radio afterglow.}

\end{deluxetable*}
\newpage
\subsection{Dark Gamma-Ray Bursts}
The need for high redshift and/or large host extinction to explain
the non-detection of a large fraction of OAs, is gradually 
disappearing. \citet{fynbo2001} and \citet{edoDARK} demonstrated that 
70\%--75\% of the OA searches conducted to date would have failed
if the GRBs were similar to that of dim bursts like GRB~000630 and 
GRB~020124 in the optical band. More recently, \citet{palli} used 
$\beta_{\mathrm{OX}}$ to introduce a new dark burst definition, 
resulting in a lower limit of the dark burst fraction of approximately 
10$\%$.
\par
With the rapid availability of small GRB error circles, first 
attainable in August 2002 via the SXC on \emph{HETE-2}, the stage was 
set for building up a small, but relatively homogeneous afterglow sample. 
This would clarify if deficient search strategies were indeed the 
reason behind the original large fraction of dark bursts. In 
Table~\ref{sxc.tab} we have listed all GRBs localized to an SXC radius 
of $\lesssim$$2\arcmin$, and where the first optical observation was
performed within $\Delta t \approx 3$\,hr.
\par
As hinted by \citet{lamb}, the majority of the SXC bursts are detected 
in the optical, or 12/14 indicating a dark burst fraction of 
10$\%$--20$\%$. The two bursts not detected in the optical are 
GRB\,020819 and GRB\,041211. We note that for nine bursts, the original 
error circle was revised a few days later, with three of those having the 
OA outside the original error circle. However, this does not affect
the dark burst statistics significantly, although it is possible
that the GRB\,041211 OA was located slightly outside the error circle.
\par
GRB\,020819 has been extensively discussed in this paper. Its optical 
faintness is mainly due to it being intrinsically faint, i.e., 
with only a modest amount of extinction the extrapolation of the 
afterglow model from the radio to the optical is consistent with our 
early $R$-band limit. The absence of an OA for GRB\,041211 is not as 
easily explained. This is predominantly due to the lack of multiwavelength 
follow-up. There is an early deep limit of $R > 21.5$\,mag at 
$\Delta t = 3.4$\,hr \citep{monfardini}, but no clear conclusions 
can be drawn without additional X-ray or radio observations.
\par
With the \emph{Swift} satellite now distributing rapid and accurate 
positions and follow-ups with the X-Ray Telescope (XRT) and 
UltraViolet/Optical Telescope (UVOT), a large homogeneous afterglow 
sample is finally being \mbox{realized}. The sample is currently rather 
small, with 23 bursts localized with the XRT (as of 2005 
May~1). In addition, the instruments are not yet fully calibrated implying 
that caution should be advised before inferring the fraction of dark 
bursts. The current optical/NIR afterglow recovery rate is 13/23, with at
least two non-detection most likely resulting from large Galactic 
extinction (GRBs\,050117 and 050421). Six of the other eight non-detections 
(GRBs\,050128, 050223, 050326, 050410, 050416B, and 050422) only 
have moderately fast and deep optical follow-ups. GRB\,050219A, 
on the other hand, was observed with 
the UVOT at $\Delta t \approx 5$~minutes to a limiting magnitude of 
$V \sim 20.7$ \citep{schady}. This burst is an excellent dark burst candidate,
although the XRT results are needed to examine its consistency with 
the fireball model. Finally, the deep optical limit for GRB\,050412 
of $R > 24.9$\,mag at $\Delta t = 2.3$\,hr \citep{kosugi} combined with
the XRT observations \citep{mangano} indicates that 
$\beta_{\mathrm{OX}} < 0.05$ for this burst, firmly establishing it as
dark.
\section{Summary \& Conclusions}
GRB\,020819, a relatively nearby burst ($z = 0.41$), is only one of
two of the 14 GRBs localized to an error radius of $\lesssim$$2\arcmin$
with the SXC on-board \emph{HETE-2}, that does not have a reported OA.
This lends support to the recent proposition that the dark burst
fraction is far lower than previously suggested, perhaps as small as 10\%.
\par
The GRB\,020819 radio afterglow is superpositioned on a faint ($R 
\approx 24.0$\,mag) blob, located around $3\arcsec$ from the center 
of a bright ($R = 19.5$\,mag) barred spiral galaxy. At $z = 0.41$ this 
corresponds to 16\,kpc. This faint blob did not show significant brightness
variations over a period of $\sim$100~days, ruling it out as the OA.
The probability that the assigned spiral host is a chance superposition 
and not physically related to the GRB, is only 0.8\%.
\par
We find that we can explain the radio afterglow without invoking large 
extinction to elucidate the absence of the OA. While we cannot rule out 
large extinction local to the GRB, it is not required. Assuming an 
average Galactic extinction curve ($R_V = 3.1$), the required host 
extinction is $A_V \approx 0.6$--1.5\,mag. 
\par
We attempted a SN search by obtaining $R$-band data at $\Delta t 
\approx 15$, 21, 27~days. We do not observe a brightening in the 
faint blob at these epochs. A 1998bw-like (2002ap-like) SN signature 
would have resulted in the blob brightening to $R \sim 22.0$\,mag
($R \sim 23.4$\,mag) for $z = 0.41$. We conclude that either the
line-of-sight is extinguished ($A_V \sim 1.5$\,mag would render a 
2002ap-like SN undetectable), GRB\,020819 is associated with a very faint 
SN, or it does not have a temporally coincident SN.
\acknowledgments
P.J. gratefully acknowledges the hospitality of Caltech, particularly
that of Shri~R. Kulkarni, Derek~B. Fox, and S.~Bradley Cenko, where 
the majority of this work was carried out. P.J. acknowledges support 
from a special grant from the Icelandic Research Council. The National 
Radio Astronomy Observatory is a facility of the National Science 
Foundation operated under cooperative agreement by Associated 
Universities, Inc. Based on observations collected at the European 
Southern Observatory Very Large Telescope at the Paranal 
Observatory under programme 69.D-0701(B). Based on observations made 
with the Nordic Optical Telescope, operated on the island of La Palma 
jointly by Denmark, Finland, Iceland, Norway, and Sweden, in the Spanish 
Observatorio del Roque de los Muchachos of the Instituto de Astrofisica 
de Canarias.



\end{document}